\documentclass[showpacs,preprintnumbers]{revtex4}
\usepackage{amssymb}
\usepackage{graphicx}
\usepackage{color}

\def\bc{\begin{center}}
\def\ec{\end{center}}

\def\beq{\begin{equation}}
\def\eeq{\end{equation}}

\begin{document}

\title{Spin Hall effect for polaritons in a TMDC
monolayer embedded in a microcavity}
\author{Oleg L. Berman$^{1,2}$, Roman Ya. Kezerashvili$^{1,2}$, and Yurii E.
Lozovik$^{3,4}$}
\affiliation{\mbox{$^{1}$Physics Department, New York
City College
of Technology, The City University of New York,} \\
Brooklyn, NY 11201, USA \\
\mbox{$^{2}$The Graduate School and University Center, The
City University of New York,} \\
New York, NY 10016, USA \\
\mbox{$^{3}$Institute of Spectroscopy, Russian Academy of Sciences,
142190 Troitsk, Moscow, Russia }\\
\mbox{$^{4}$MIEM at National Research University Higher School of
Economics, Moscow, Russia}}
\date{\today}

\begin{abstract}
The spin Hall effect for polaritons (SHEP) in a transition metal
dichalcogenides (TMDC) monolayer embedded in a microcavity is
predicted.  We demonstrate that two counterpropagating laser beams
incident on a TMDC monolayer can deflect a superfluid polariton flow
due to the generation the effective gauge vector and scalar
potentials. The components of polariton conductivity tensor for both
non-interacting polaritons without Bose-Einstein condensation (BEC)
and for weakly-interacting Bose gas of polaritons in the presence of
BEC and superfluidity are obtained.
 It is shown that the polariton flows in the same valley
are splitting: the superfluid components of the \textit{A} and
\textit{B} polariton flows propagate in opposite directions along
the counterpropagating beams, while the normal components of the
flows slightly deflect in opposite directions and propagate almost
perpendicularly to the beams. The possible experimental observation
of SHEP in a microcavity is proposed.

\end{abstract}

\maketitle


\section{Introduction}

\label{intro}

Transition metal dichalcogenides (TMDCs)  monolayers such as $%
\mathrm{MoS_{2}}$, $\mathrm{WS_{2}}$, $\mathrm{MoSe_{2}}$, $\mathrm{WSe_{2}}$%
, $\mathrm{MoTe_{2}}$, and $\mathrm{WTe_{2}}$ are characterized by
the direct gap in a single-particle spectrum exhibiting the
semiconducting band structure and strong spin-orbit
coupling~\cite{Kormanyos,Mak2010,Mak2012,Novoselov,Ross,Zhao,Glazov}.
Significant spin-orbit splitting in the valence
band leads to the formation of two distinct  types \textit{A} and \textit{B}~excitons \cite%
{Reichman}.  \textit{A} excitons are formed by spin-up electrons
from conduction and spin-down holes from valence band, while type
\textit{B} excitons are formed by spin-down electrons from
conduction and spin-up holes from valence band~\cite{Glazov}. When
excitons are created optically, the optical field couples only to
the orbital part of the wave function, while the spin is conserved
in optical transitions~\cite{Xiao}. In the conduction and valence
bands of TMDC the electron wave function is given by the linear
superposition of $p$ and $d$ orbitals due to the orbital
hybridization~\cite{Yakobson}. Caused by coupling to the optical
field, the total angular momentum change between $p$ and $d$
orbitals is $1$. The corresponding change of the total moment is
compensated by the photon spin $1$ due to conservation of the total
angular momentum~\cite{Yakobson}.

Recently, microcavity polaritons, formed by excitons in TMDCs
embedded in a microcavity, attracted the interest of experimental
and theoretical studies. The exciton polaritons, formed by cavity
photons and excitons in $\mathrm{MoS_{2}}$~\cite{Menon} and
$\mathrm{WS_{2}}$~\cite{Smith} monolayers, and a $\mathrm{MoSe_{2}}$
monolayer supported by $h$-BN layers~\cite{Tartakovskii}, embedded
in a microcavity, were observed experimentally at room temperature.
The exciton polariton modes
formed due to interaction coupling of excitons in $\mathrm{MoS_{2}}$ and $%
\mathrm{WS_{2}}$ monolayers and microcavity photons were
studied~\cite{Kavokin_PRB_2015}. In Ref. \cite{Kavokin_2D_2017} the
phase diagram of polariton Bose-Einstein condensation (BEC) in a
microcavity with an embedded $\mathrm{MoS_{2}}$ monolayer was
presented. An experimentally relevant range of parameters, at which
room-temperature superfluidity of exciton polaritons can be observed
in a microcavity with an embedded $\mathrm{MoS_{2}}$ monolayer, was
determined in Ref.~\cite{Pomirchi} in the framework of driven
diffusive dynamics, while in Ref.~\cite{Korean} the phase diagram of
polariton condensate, formed by TMDC excitons coupled to microcavity
photons, was studied theoretically.

Strong spin-orbit coupling in TMDCs can lead to the spin Hall effect
(SHE), which is one of the most essential effects in spintronics~\cite%
{Das_Sarma_review,Eschrig_review}. The SHE is the result of
generation of a transverse spin current as a response to a
longitudinal applied electric field that results in spin
accumulation of the carriers with opposite spins at the opposite
edges of  samples~\cite{Perel,Hirsh,Zhang}. Under applied electric
field this transverse spin current can be generated in the systems
with strong SOC due either the properties of the electron band
structure or scattering on the impurities~\cite{Perel,Hirsh,Zhang}.

A method to observe the SHE for cold atoms under light-induced
gauge potentials was proposed~in Refs. \cite{Ohrberg,Klein,jpb}. This gauge
potential is created when two coordinate dependent laser beams interact with
three-level atoms. The vector potential leads to a nonvanishing effective
magnetic field, if at least one of the two light beams has a vortex~\cite%
{Ohrberg,Klein}. A nonvanishing effective magnetic field can be
created without existence of a vortex in a laser beam, if two
counterpropagating and overlapping laser beams with shifted spatial
profiles interact with three-level atoms~\cite{pra}. The spin Hall
effect for cold atoms can be observed when two counterpropagating
Gaussian laser beams with shifted centers generate a spatially
slowly varying gauge field acting on three-level atoms~\cite{Duan}.

The spin Hall effect for excitons (SHEE)  in TMDC was proposed for
circularly polarized pumping in Ref.~\cite{li}, where the mechanism
is based on creation of the gauge vector and scalar potentials due
to the coupling of excitons to two counterpropagating and
overlapping laser beams. The exciton Hall effect (EHE) in TMDC was
studied in Ref.~\cite{Onga}, where the exciton flows deflect due to
the peculiarities of internal structure of TMDC for linearly and
circularly polarized pumping. The optical spin Hall effect for
microcavity polaritons (OSHEP), formed by excitons in a GaAs quantum
well, embedded in a high-quality GaAs/AlGaAs microcavity, caused by
the polarization dependence of Rayleigh scattering of light by
structural disorder in microcavities, was studied for linearly
polarized pumping~\cite{OSH1,OSH2}.

\begin{figure}[t]
\centering
\includegraphics[width=7.5cm]{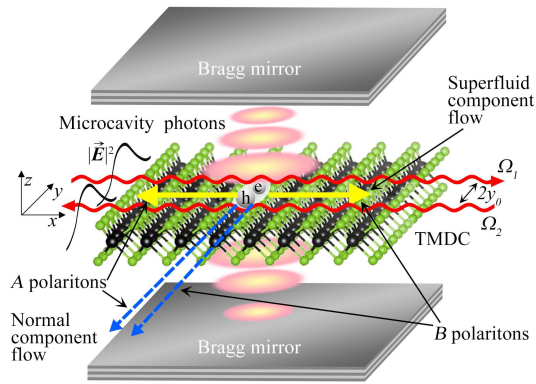}
\caption{(Color online) Schematic representation of the
light-induced  spin Hall effect for polaritons in a TMDC monolayer
embedded in a microcavity. The solid and dashed lines with arrows
show the directions of deflected superfluid and normal flows of A
and B polaritons, correspondingly. Two counterpropagating laser
beams are shown by waved lines. } \label{Fig1}
\end{figure}

In this paper we predict the spin Hall effect for polaritons (SHEP),
formed by microcavity photons and excitons in TMDC materials
embedded into a microcavity.  The schematic representation of the
light-induced spin Hall effect for microcavity polaritons in a TMDC
layer is depicted in Fig.~\ref{Fig1} and can be described as
follows. Two Bragg mirrors placed opposite each other at the
antinodes of the confined photonic mode form a microcavity, and a
TMDC layer  is embedded parallel to the Brag mirrors within the
cavity. As a result of the laser pumping the resonant exciton-photon
interaction leads to the Rabi splitting in the excitation spectrum~\cite%
{Carusotto_rmp,Snoke_book}. The polaritons cloud is formed due to
the coupling of excitons created in a TMDC layer and microcavity
photons. The mechanism of the SHEP is following. Two
coordinate-dependent, counterpropagating and overlapping laser beams
in the plane of the TMDC layer interact with a cloud of polaritons.
These laser beams, characterized by Rabi frequencies $\Omega _{1}$
and $\Omega _{2}$ produce the spin-dependent gauge magnetic and
electric fields~\cite{pra,Duan} due to strong SOC for blue electrons
and holes in TMDC~\cite{li}. Excitons forming polaritons in these
gauge magnetic and electric fields form spin-dependent light dressed
states~\cite{Shirley,Zeldovich,Sambe}
due to the interaction with laser beams. Below we show that the
gauge magnetic field splits the \textit{A} and \textit{B} polariton
flows. The normal components of the \textit{A} and \textit{B}
polariton flows slightly deflect in opposite directions and
propagate almost perpendicularly to the counterpropagating beams. In
contrast, the superfluid components of the \textit{A} and \textit{B}
polariton flows propagate in opposite directions along the
counterpropagating beams. Therefore, one can observe the
light-induced spin Hall effect for microcavity polaritons, formed by
excitons in a TMDC layer. For the laser pumping frequencies,
corresponding to the resonant excitations of one type of excitons
(\textit{A} or \textit{B}), the corresponding excitons together with
coupled to them photons form polaritons, which deflect to only one
 direction along the counterpropagating beams.  The flow
of polaritons, associated with this spin current, results in the
flow of photons, coupled to excitons in a TMDC layer. Therefore, we
propose the method to control photon flows. We are considering the
SHEP in two regimes: non-interacting polaritons with the quadratic
spectrum in a very dilute limit, when the polariton density is not
enough to create BEC at a given temperature and the limit of higher
polariton densities in the presence of BEC and superfluidity. Also
we propose the method to experimentally observe the superfluidity of
microcavity polaritons due to the spin Hall flow of polaritons.

 For our study we assume that the pumping beam is
circularly polarized, and hence the polaritons are formed by
excitons only in one of the valleys: $\mathbf{K}$ or
$\mathbf{-K}$~\cite{Xiao,Mak2013}. Below we focus on formation of
polaritons in $\mathbf{K}$ valley and an extension to $\mathbf{-K}$
valley is obvious.

The paper is organized in the following way. In Sec.~\ref{eff_pol},
we present the effective Hamiltonian for microcavity polaritons,
formed by cavity photons and TMDC excitons, coupled to two laser
beams, which is producing the gauge vector and scalar potentials.
The tensor of the polariton conductivity, which is the linear
response of the polariton flow on the scalar gauge field, and the
corresponding resistivity tensor for non-interacting
microcavity polaritons in the SHEP regime are obtained in Sec.~\ref{SHEideal}%
. The conductivity tensor for microcavity polaritons in the presence
of superfluidity in the SHEP regime is derived in Sec.~\ref{SHEBEC}.
 In Sec.~\ref{exper}, we discuss the possibility to
observe the SHEP. The technological applications of the SHEP in TMDC
monolayers are considered in Sec.~\ref{tech}. The discussion of our
results is presented in Sec.~\ref{disc}. Conclusions follow in
Sec.~\ref{conc}.

\section{Microcavity polaritons in the presence of counterpropagating laser beams}

\label{eff_pol}

Let us consider the effective Hamiltonian of polaritons, formed by
TMDC excitons coupled to microcavity photons in the presence of
counterpropagating and overlapping laser beams. The deflection of
polaritons occurs via the action of the laser beams on the exciton
component of the polaritons. The exciton component of polaritons in
a TMDC heterostructure is coupled to two counterpropagating and
overlapping coordinate dependent infrared laser beams. The coupling
of TMDC excitons to two coordinate dependent laser beams, moving
along the plane of
 TMDC, results in the gauge vector and scalar potentials {\cite{pra,Duan}%
}, acting on the centers-of-mass of TMDC excitons~\cite{li}. The
latter causes the spin Hall effect for excitons (SHEE) in TMDC.

The Hamiltonian of TMDC polaritons in the presence of
counterpropagating and overlapping laser beams, can be written as

\begin{equation}
\hat{\mathcal{H}}=\hat{H}_{exc}+\hat{H}_{ph}+\hat{H}_{exc-ph}+\hat{H}_{exc-exc},
\label{Ham_pol_tot}
\end{equation}%
where $\hat{H}_{exc}$ is the Hamiltonian of excitons in the gauge field
produced by two counterpropagating and overlapping laser beams, $\hat{H}%
_{ph} $ is the Hamiltonian of microcavity photons, $\hat{H}_{exc-ph}$ is the
Hamiltonian of exciton-photon coupling, and $\hat{H}_{exc-exc}$ is the
Hamiltonian of exciton-exciton interaction.

The Hamiltonian of $2D$ excitons in the presence of
counterpropagating and overlapping laser beams can be presented as
\begin{equation}
\hat{H}_{exc}=\sum_{\mathbf{P}}^{{}}\varepsilon _{ex}(\mathbf{P})\hat{b}_{%
\mathbf{P}}^{\dagger }\hat{b}_{\mathbf{P}},  \label{Ham_exc}
\end{equation}
where $\hat{b}_{\mathbf{P}}^{\dagger }$ and $\hat{b}_{\mathbf{P}}$ are
excitonic Bose creation and annihilation operators obeying Bose commutation
relations. In Eq.~(\ref{Ham_exc}), $\varepsilon _{ex}(\mathbf{P}%
)=E_{bg}-E_{b}+\varepsilon _{0}(\mathbf{P})$ is the energy dispersion of a
single exciton in a TMDC layer, where $E_{bg}$ is the band gap energy, $%
E_{b} $ is the binding energy of an exciton, and $\varepsilon _{0}(\mathbf{%
P})$ is the energy spectrum of a single exciton in a TMDC coupled to
two infrared, coordinate dependent laser beams. The interaction of
the exciton in a TMDC monolayer with two counterpropagating Gaussian
laser beams can be represented as the interaction with the gauge
vector and  scalar potentials~\cite{li}. We analyze the Hamiltonian
for non-interacting TMDC excitons coupled to two counterpropagating
Gaussian laser beams in Appendix~\ref{app:C}. By expanding the gauge
vector potential, acting on excitons, in the linear order with
respect to the coordinate, the constant gauge magnetic field is
obtained. The gauge scalar potential, acting on excitons, is the
even function of the coordinate and, therefore, has no linear order
term with respect to the coordinate, and, therefore,
can be omitted. In this case, $\varepsilon _{0}(\mathbf{%
P})$ can be written as~\cite{li}
\begin{equation}
\label{e0} \varepsilon _{0}(\mathbf{P})= \frac{\left(
\mathbf{P}-\mathbf{A}_{\sigma }\right)^{2}}{2M} ,
\end{equation}
 where $M$ is the mass of an exciton and
$\mathbf{A}_{\sigma }$ is the gauge vector potential acting on the
exciton component of polaritons, associated with
different spin states of the conduction band electron, forming an exciton, $%
\sigma =\uparrow $ and $\downarrow$.

The Hamiltonian of non-interacting photons in a microcavity has the
form~\cite{Pau}:
\begin{equation}
\hat{H}_{ph}=\sum_{\mathbf{P}}\varepsilon _{ph}(P)\hat{a}_{\mathbf{P}%
}^{\dagger }\hat{a}_{\mathbf{P}},  \label{Ham_ph}
\end{equation}
where $\hat{a}_{\mathbf{P}}^{\dagger }$ and $\hat{a}_{\mathbf{P}}$ are
photonic creation and annihilation Bose operators. In Eq.~(\ref{Ham_ph}) $%
\varepsilon _{ph}(P)=(c/\tilde{n})\sqrt{P^{2}+\hbar ^{2}\pi
^{2}q^{2}L_{C}^{-2}}$ is the spectrum of the microcavity photons,
where $c$ is the speed of light, $L_{C}$ is the length of the cavity, $%
\tilde{n}=\sqrt{\epsilon }$ is the effective index of refraction of
the microcavity, $\epsilon $ is the dielectric constant of the
cavity, and $q$ is the integer, which represents the longitudinal
mode number.

The Hamiltonian of harmonic exciton-photon coupling is given
by~\cite{Ciuti}:
\begin{equation}
\hat{H}_{exc-ph} = {\hbar \Omega_{R}}\sum_{\mathbf{P}} \hat{a}_{\mathbf{P}%
}^{\dagger}\hat{b}_{\mathbf{P}}^{} + h.c. .  \label{Ham_exph}
\end{equation}
In Eq.~(\ref{Ham_exph}) $\Omega _{R}$ is the Rabi splitting constant
which represents the exciton-photon coupling energy and is defined
by  the dipole matrix element,
corresponding to the transition with the exciton formation~\cite%
{Savona_review} and has different values depending on the material where
polaritons are formed.

Below we consider two regimes: i) a very dilute system of
non-interacting polaritons when the exciton-exciton interaction is
neglected, i. e. $H_{exc-exc} = 0$ and ii) a weakly interacting
Bose-gas of polaritons characterized by superfluidity, when
exciton-exciton hard core repulsion is taken into account. In this
Section we focus on the regime (i), while the regime (ii) is
discussed in Sec.~\ref{SHEBEC}. Assuming $\hat{H}_{exc-exc}=0$ the
Hamiltonian $\hat{\mathcal{H}}$ can be diagonalized by using
 the unitary
transformation as presented in Appendix~\ref{app:B}. Substituting Eq.~(\ref%
{bog_tr}) into~(\ref{Ham_pol_tot}), one obtains the Hamiltonian of
lower polaritons~\cite{Ciuti}:
\begin{equation}
\hat{\mathcal{H}}_{0} =\sum_{\mathbf{P}}\varepsilon _{LP}(\mathbf{P})\hat{p}_{\mathbf{P}%
}^{\dagger }\hat{p}_{\mathbf{P}},  \label{Ham_tot_p}
\end{equation}%
where $\hat{p}_{\mathbf{P}}^{\dagger }$ and $\hat{p}_{\mathbf{P}}$
are the Bose creation and annihilation operators for the lower
polaritons. The single-particle lower polariton spectrum,  which one
obtains from Eq.~(\ref{eps0}) by substituting~(\ref{e0}) and  the
expression for the spectrum of the microcavity photons, is given by
\begin{equation}
\varepsilon _{LP}(\mathbf{P})=\hbar \pi qL_{C}^{-1}-|\hbar \Omega
_{R}|+\varepsilon (\mathbf{P}),  \label{eps00}
\end{equation}%
where $\varepsilon (\mathbf{P})$ has the form
\begin{equation}
\varepsilon(\mathbf{P}) = \frac{1}{2}\left(\varepsilon_{0}(\mathbf{P})  + \frac{P^{2}}{2m_{ph}}\right) = \frac{1}{2}\left(\frac{\left(\mathbf{P} - \mathbf{A%
}_{\sigma}\right)^{2}}{2M} + \frac{P^{2}}{2m_{ph}}\right).
\label{ex_phot}
\end{equation}
In Eq.~(\ref{ex_phot}) $m_{ph}=\hbar \pi q/((c/\tilde{n})L_{C})$ is
the effective mass of microcavity photons, and it is obtained under
the assumption of small momenta in
the first order with respect to the small parameter $\alpha \equiv 1/2(M^{-1}+(c/%
\tilde{n})L_{C}/\hbar \pi q)P^{2}/|\hbar \Omega _{R}|\ll 1$.

After simple algebraic transformation Eq.~(\ref{ex_phot}) can be
rewritten as
\begin{equation}
\varepsilon (\mathbf{P})=\frac{\left( \mathbf{P}-\mathbf{A}_{\sigma
}^{(eff)}\right) ^{2}}{2M_{p}}+V^{(eff)},  \label{ex_phot3}
\end{equation}
%
where $M_{p}=2\mu $, $\mu =Mm_{ph}/\left( M+m_{ph}\right) $ is the
exciton-photon reduced mass. In Eq.~(\ref{ex_phot3})
$\mathbf{A}_{\sigma }^{(eff)}$ and $V^{(eff)}$ are the effective
vector and scalar potentials, respectively, acting on polaritons,
and are given by
\begin{equation}
\mathbf{A}_{\sigma }^{(eff)}=\frac{m_{ph}\mathbf{A}_{\sigma }}{M+m_{ph}},%
\text{ \ \ \ \ }V^{(eff)}=\frac{A_{\sigma }^{2}}{4\left( M+m_{ph}\right) }.
\label{eff_scalar_vec}
\end{equation}
%
In Eq.~(\ref{eff_scalar_vec}) $\mathbf{A}_{\sigma }$ is the gauge
vector potential acting on excitons, obtained in Ref.~\cite{li}  and
given by Eq.~(\ref{ABexp}). Let us mention that $\mathbf{A}_{\sigma
}^{(eff)}$ and $V^{(eff)}$ are obtained employing diagonalization of
the Hamiltonian $\hat{\mathcal{H}}$ under assumption
$\hat{H}_{exc-exc}=0$ by using the unitary transformation presented
in Appendix~\ref{app:B}. It follows from Eq.~(\ref{eff_scalar_vec})
that acting on polaritons effective gauge scalar potential
$V^{(eff)}$, determined by the gauge vector potential
$\mathbf{A}_{\sigma }$ acting on excitons, depends on $y$, while the
gauge scalar potential $V_{\sigma}$ acting on excitons in the linear
order with respect to $y$ is a constant. Therefore, $V^{(eff)}$
leads to non-zero effective gauge electric field
$\mathbf{E}^{(eff)}$ acting on polaritons. In contrast, the gauge
electric field acting on excitons is zero in the linear order with
respect to $y$.

Substituting Eq.~(\ref{ABexp}) into (\ref{eff_scalar_vec}),
assuming slowly changing gauge potential, and keeping only terms,
linear with respect to $y$, the effective gauge vector and scalar
potentials acting on the polaritons can be written as
\begin{equation}
\mathbf{A}_{\sigma }^{(eff)}=\frac{m_{ph}\eta _{\sigma }\hbar \left(
|k_{1}|+|k_{2}|\right) }{2\left( M+m_{ph}\right) }\left( 1+\frac{y}{2l}%
\right) \mathbf{e}_{x},\text{ \ }V^{(eff)}=\frac{\hbar ^{2}\left(
|k_{1}|+|k_{2}|\right) ^{2}}{16\left( M+m_{ph}\right) }\left( 1+\frac{y}{l}%
\right) .  \label{Vec_slalar_liniar}
\end{equation}
Therefore, the effective gauge magnetic  $\mathbf{B}_{\sigma
}^{(eff)}$  and  electric fields $\mathbf{E}^{(eff)}$ fields acting
along $z$- and $y$-axis, respectively, are
\begin{equation}
\mathbf{B}_{\sigma }^{(eff)}=\nabla _{\mathbf{R}}\times \mathbf{A}_{\sigma
}^{(eff)}=\frac{-\eta _{\sigma }\hbar m_{ph}\left( |k_{1}|+|k_{2}|\right) }{%
4l\left( M+m_{ph}\right) }\mathbf{e}_{z},\text{ \ }\mathbf{E}%
^{(eff)}=-\nabla _{\mathbf{R}}V^{(eff)}=-\frac{\hbar ^{2}\left(
|k_{1}|+|k_{2}|\right) ^{2}}{16l\left( M+m_{ph}\right) }\mathbf{e}_{y}.
\label{Beff_Aeff}
\end{equation}

The analysis of Eqs.~(\ref{Vec_slalar_liniar}) and~(\ref{Beff_Aeff})
shows that the effective gauge vector potential and effective
magnetic field are different for \textit{A} and \textit{B}
polaritons due to the factor $\eta_{\uparrow} = 1$  for an
\textit{A} exciton and $\eta_{\downarrow} = -1$ for a \textit{B}
exciton, while   the effective gauge scalar potential and effective
electric field do not depend on the spin orientation $\sigma$. As a
result, the effective gauge magnetic field $\mathbf{B}_{\sigma
}^{(eff)}$ deflects the  polaritons consisting from the excitons
with different spin states of charge carriers (\textit{A} and
\textit{B} excitons) towards opposite directions. Thus, the system
under consideration demonstrates the SHEP. Also as mentioned above,
there is an effective uniform electric field acting on polaritons,
in contrast to absence of such uniform field for excitons.

Let us mention that the SHEE was proposed for interlayer excitons in
a $\mathrm{MoSe_{2}-WSe_{2}}$ van der Waals heterostructure, where
electrons are located in a $\mathrm{MoSe_{2}}$
monolayer, while holes are located in a $\mathrm{WSe_{2}}$ monolayer~\cite%
{li}. The interlayer excitons in such heterostructures are
characterized by relatively high lifetime, due to suppression of the
electron-hole recombination since electrons and holes are spatially
separated in different monolayers~\cite{Lozovik,LozRep}. However,
the increase of the exciton lifetime does not essentially influence
the lifetime of polaritons, because the polariton lifetime is
determined by the lifetime of the microcavity photons. The lifetime
of the microcavity photons is much smaller than the lifetime of
excitons, since the microcavity photons leave the microcavity much
faster than electrons and hole recombine. Therefore, one can
consider the excitons in a single TMDC monolayer, embedded in a
microcavity, without sufficient decrease of the polariton lifetime
compared with the TMDC van der Waals heterostructure.

\section{Resistivity and conductivity tensors for non-interacting
microcavity polaritons in the SHEP regime}

\label{SHEideal}

In this Section, we consider the dilute system of non-interacting
microcavity polaritons when the concentration $n$ is too low to form
the BEC at given temperature. Applying the Drude model, one can
write the transport
equation for microcavity polaritons, moving in both the effective electric $%
\mathbf{E}^{(eff)}$ and magnetic $\mathbf{B}_{\sigma }^{(eff)}$ fields as~%
\cite{Simon,Snoke_book_1}
\begin{equation}
\frac{d\mathbf{P}}{dt}=\mathbf{E}^{(eff)}+\mathbf{v}\times \mathbf{B}%
_{\sigma }^{(eff)}-\frac{\mathbf{P}}{\tau },  \label{tran}
\end{equation}
%
where $\mathbf{v}$ is the velocity and $\tau $ is a scattering
time of microcavity polaritons. For a steady state, setting $d\mathbf{P}%
/dt=0 $, and using $\mathbf{P}=M_{p}\mathbf{v}$, one obtains
\begin{equation}
\mathbf{E}^{(eff)}=\frac{M_{p}}{n\tau }\mathbf{j}-\frac{\mathbf{j}\times
\mathbf{B}_{\sigma }^{(eff)}}{n},  \label{Eeq}
\end{equation}
where the linear polariton flow density is defined as $\mathbf{j}=n%
\mathbf{v}$.

Following Ref.~\cite{Snoke_book_1}, for the resistivity and
conductivity tensors for polaritons moving in effective electric and
magnetic fields we use the similar definitions. In particular, the
 $2\times 2$ resistivity matrix $\varrho _{\sigma }$ can be defined as $%
\mathbf{E}^{(eff)}=\varrho _{\sigma }\mathbf{j},$ with the diagonal
$\rho _{\sigma xx}$ and $\rho _{\sigma yy},$ and the off-diagonal
$\rho _{\sigma xy}$ and $\rho _{\sigma yx}$ components known as the
Hall resistivity \cite{Simon} given by
\begin{equation}
\rho _{\sigma xx}=\rho _{\sigma yy}=\frac{M_{p}}{n\tau },\text{ \ \ }\rho
_{\sigma xy}=-\rho _{\sigma yx}=\frac{\eta _{\sigma }B^{(eff)}}{n},
\label{resistivity}
\end{equation}
%
where $B^{(eff)}$ is the magnitude of the effective magnetic field $\mathbf{B%
}_{\sigma }^{(eff)}$.

We define the Hall coefficient $R_{H\sigma }$ as
\begin{equation}
R_{H\sigma }=\frac{\rho _{\sigma yx}}{B^{(eff)}}=-\frac{\eta _{\sigma }}{n}.
\label{Hall}
\end{equation}%

The conductivity tensor $\tilde{\sigma}_{\sigma }$ is defined as the
inverse matrix to the resistivity matrix $\varrho _{\sigma }$. In
our case of the Hall conductivity, the diagonal and off-diagonal
components of $\tilde{\sigma}_{\sigma }$  are given by
\begin{equation}
\sigma _{\sigma xx}=\sigma _{\sigma yy}=\frac{\sigma _{0}}{1+\omega
_{c}^{2}\tau ^{2}},\text{ \ }\sigma _{\sigma xy}=-\sigma _{\sigma
yx}= - \frac{\eta _{\sigma }\sigma _{0}\omega _{c}\tau }{1+\omega
_{c}^{2}\tau ^{2}},  \label{conductivity}
\end{equation}
where $\sigma _{0}=\tau n/M_{p}$ and $\omega _{c}=B^{(eff)}/M_{p}$
is the cyclotron frequency.  As it can be seen from
Eqs.~(\ref{resistivity}),~(\ref{Hall}), and~(\ref{conductivity}),
the Hall resistivity, Hall coefficient, and Hall conductivity depend
on the spin orientation $\sigma$.

The $x$ and $y$ components of the linear polariton flow density are
defined as $j_{x} = - \sigma_{\sigma xy} E^{(eff)}$ and $j_{y} = -
\sigma_{\sigma yy} E^{(eff)}$.

\begin{figure}[h]
\centering
\includegraphics[width=7.0cm]{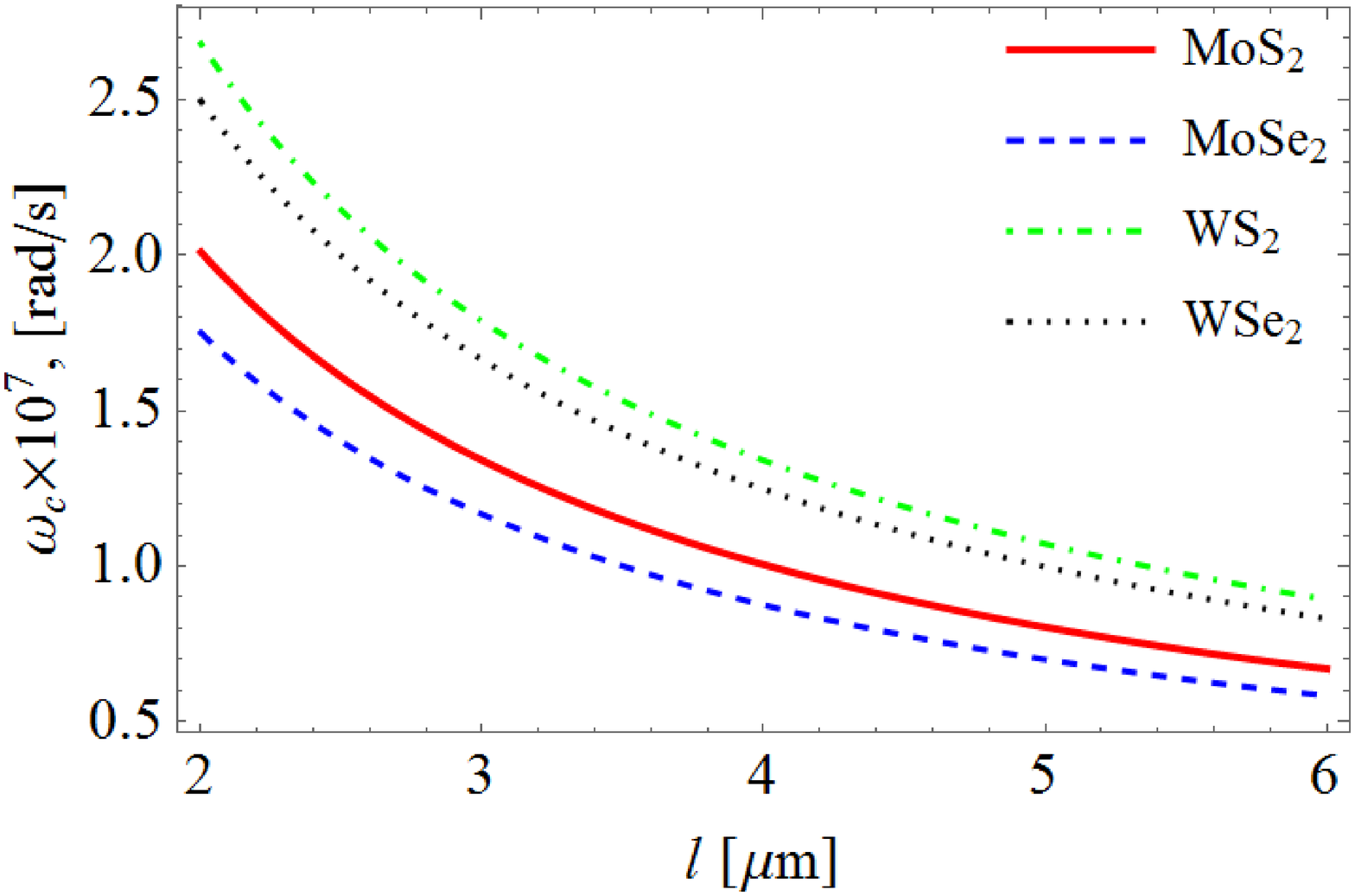}
\includegraphics[width=7.0cm]{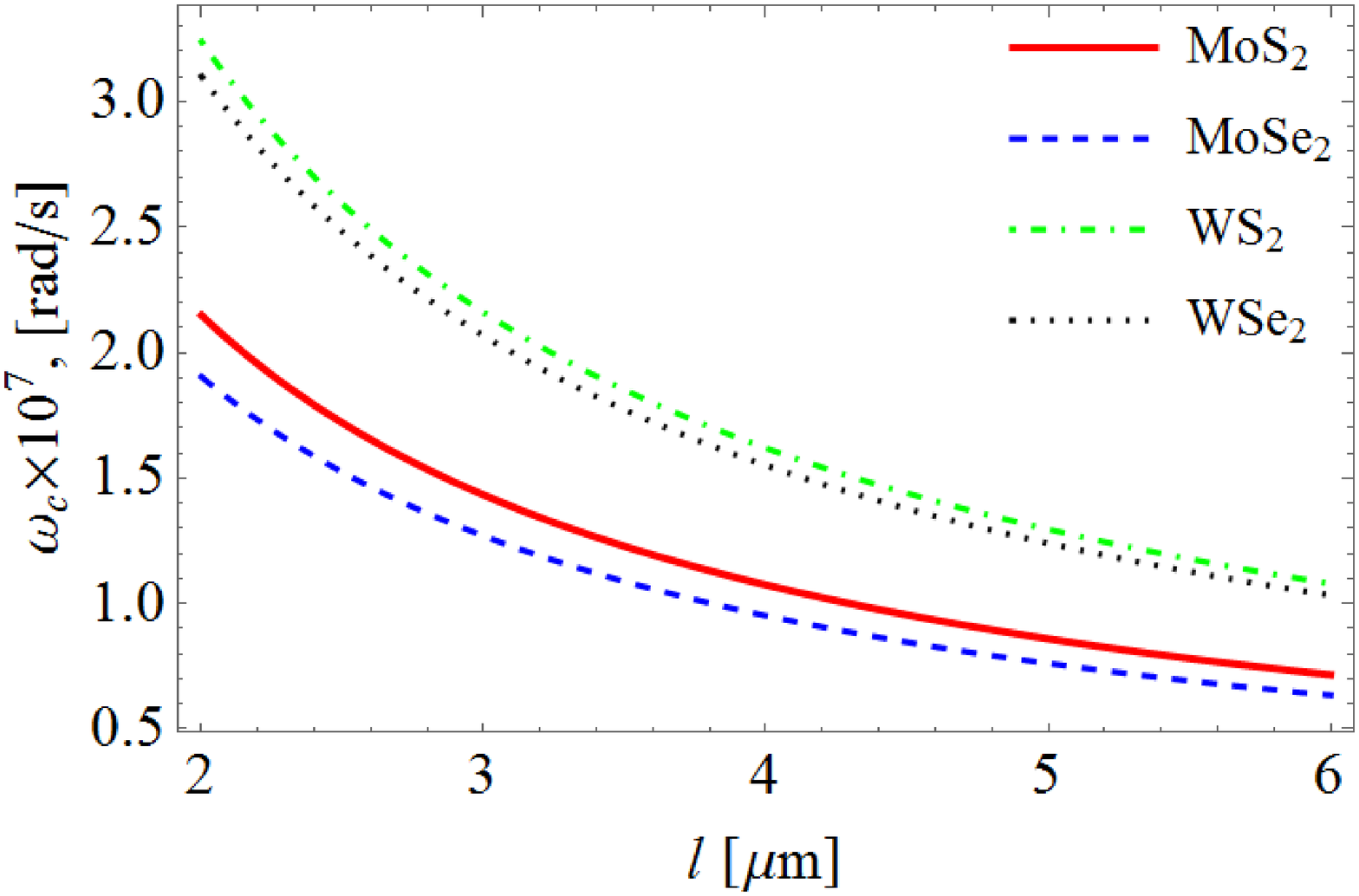}
\caption{(Color online) The dependence of the cyclotron frequency
$\omega _{c}$  on  the parameter $l$, determined by the distance
between the counterpropagating laser beams. Calculations performed
for $|k_{1}|+|k_{2}| = 3 \ \mathrm{\mu m}^{-1}$. (a) and (b)
$\omega_{c}$ as a function of $l$ for \textit{A} and \textit{B}
polaritons, respectively.}
 \label{omega_fig}
\end{figure}

The dependencies of  the cyclotron frequency $\omega_{c}$ on the
distance between the centers of the contrpropagated laser beams for
\textit{A} and \textit{B} polaritons are presented in
Fig.~\ref{omega_fig}. According to Fig.~\ref{omega_fig}, one can
conclude that the cyclotron frequency $\omega _{c}$  decreases  with
the parameter $l$, and it is the largest for a $\mathrm{WS_{2}}$
monolayer and the smallest for a $\mathrm{MoSe_{2}}$  monolayer at
the same $l$. Also, for the same TMDC monolayer $\omega _{c}$ is
larger for \textit{B} polaritons than for \textit{A} polaritons.

 In our calculations and estimations we use the following parameters.
For the microcavity we use the parameters
from Ref.~\cite{Tartakovskii}: $\tilde{n}=2.2$, $%
L_{C}=2.3\ \mathrm{\mu m}$, $q=5$ and obtain the effective mass of
microcavity photons: $m_{ph}=5.802\times 10^{-6}m_{0}$, where
$m_{0}$ is the mass of an electron. The experimentally measured
values of the Rabi splitting constant $\hbar \Omega _{R}$ were
reported as $46\pm 3\ \mathrm{meV}$~\cite{Menon}, $20\ \mathrm{meV}$~\cite%
{Tartakovskii}, and $70\ \mathrm{meV}$~\cite{Smith} for
$\mathrm{MoS_{2}}$, $\mathrm{MoSe_{2}}$, and $\mathrm{WS_{2}}$
monolayers, respectively.  We use the sets of effective masses for
electrons and holes in various TMDCs from Refs.~\cite{emass,hmass}.

 The scattering time of microcavity polaritons is $\tau (P) = X_{\mathbf{P}%
}^{-4} \tau_{ex}(P) \approx 4 \tau_{ex}(P)$~\cite{BKL}, where
$X_{\mathbf{P}}$ is defined by Eq.~(\ref{bog}). The exciton
relaxation time $\tau_{ex}(P)$ can be approximated by its average value $%
\bar{\tau}_{ex} = \langle \tau_{ex}(P)\rangle$, which can be
obtained from the exciton mobility $\tilde{\mu}_{ex} = e
\bar{\tau}_{ex}/M$. For a WSe$_{2}$ monolayer the exciton relaxation
time was measured as $\bar{\tau}_{ex} = 260 \
\mathrm{fs}$~\cite{Cadiz}. Therefore, a scattering time of
microcavity polaritons can be estimated as $\tau = 4 \bar{\tau}_{ex}
 = 1.040 \ \mathrm{ps}$, which is less than the
polariton lifetime estimated as tens of
picoseconds~\cite{Snoke_book}.  In our calculations we use this
value for the scattering time of microcavity polaritons $\tau$.

According to Ref.~\cite{li}, for the SHEE the cyclotron frequency is
$\omega_{c}\sim 1/M$, where $M$ is the mass of the exciton. For the
SHEP According to Eq.~(\ref{Beff_Aeff}) for polaritons $B^{(eff)}
\sim m_{ph}/\left(M+m_{ph}\right)$. Since $m_{ph} \ll M$ we have
$M_{p} \approx m_{ph}$, and one concludes that for the  SHEP
cyclotron frequency is also $\omega_{c} \sim 1/M$. Therefore, the
cyclotron frequencies for the SHEP and the SHEE will be almost the
same order of magnitude, and the degree to which spin currents are
separated in the case of exciton-polaritons will be approximately
the same as in the regular exciton case.

%
%
%
%
%


\section{Spin Hall effect for microcavity polaritons in the presence of
superfluidity}

\label{SHEBEC}

In the dilute limit $na_{2D}^{2}\ll 1$, where $a_{2D}$ is the 2D
exciton Bohr radius, at sufficiently low temperatures the
Bose-Einstein condensation of polaritons appears in the system, and
the corresponding Hamiltonian of a weakly-interacting Bose gas of
microcavity polaritons with hard-core repulsion is presented, for
example, in Ref.~\cite{BLS}. For the simplicity one can consider the
Thomas-Fermi approximation for the polariton
condensate density profile in the non-quantizing effective magnetic field $%
\mathbf{B}_{\sigma }^{(eff)}$. Within this approximation the
polariton condensate density profile for the system does not depend
on the effective magnetic field. The Thomas-Fermi approximation is
valid if the healing length $\xi $~\cite{Pitaevskii} of polaritons
is much less than the other characteristic length parameters of the
system such as the effective magnetic length:  $\xi \ll
r_{B}^{(eff)}$. The healing length for polaritons $\xi $ is given by
$\xi =\hbar /\sqrt{2M_{p}\mu _{p}}$, where $\mu _{p}$ is the
chemical potential of weakly-interacting Bose gas of polaritons in
the Bogoliubov approximation~\cite{Abrikosov,Griffin}
$\mu_{p}=U_{\mathrm{eff}}^{(0)}n$, where $U_{\mathrm{eff}}^{(0)}$ is
the Fourier transform of the effective polariton-polariton pair
repulsion potential, given by the hard-core contact
potential. In Ref. \cite{BLS} $U_{\mathrm{eff}}^{(0)}$ is defined as $U_{%
\mathrm{eff}}^{(0)}=3ke^{2}a_{2D}/(2\epsilon )$, where $k=9\times
10^{9}\
\mathrm{N \cdot m^{2}/C^{2}}$, $\epsilon =\tilde{n}%
^{2}$ is the dielectric constant of the microcavity, and
$a_{2D}=\hbar ^{2}\epsilon /\left( 2\mu _{ex}ke^{2}\right) $ is the
2D exciton Bohr radius, and $\mu_{ex}$ is the exciton reduced mass,
defined as $\mu_{ex}=m_{e\uparrow }m_{h\downarrow }/\left(
m_{e\uparrow }+m_{h\downarrow }\right) $ and
$\mu_{ex}=m_{e\downarrow }m_{h\uparrow }/\left( m_{e\downarrow
}+m_{h\uparrow }\right) $ for \textit{A} and \textit{B} excitons,
correspondingly.

In experiments, the exciton density in a TMDC monolayer was obtained up to $%
n = 5 \times 10^{11}\ \mathrm{cm^{-2}}$~\cite{You_bi}, and we use
this value for $n$ in our estimations. Since 2D  Bohr radius is
$a_{2DA}= 3.875 \ \mathrm{{\mathring{A}}}$ and $a_{2DB}= 4.41 \
\mathrm{{\mathring{A}}}$  for \textit{A} and \textit{B} excitons,
respectively, one correspondingly obtains $n a_{2DA}^{2} = 7.509
\times 10^{-4}\ll 1$ and $n a_{2DB}^{2} = 9.723 \times 10^{-4}\ll
1$. Therefore, the polariton system can be treated as a
weakly-interacting Bose gas. We
estimate the healing length as $\xi_{A} = 1.947 \ \mathrm{\mu m}$ and $%
\xi_{B} = 1.825 \ \mathrm{\mu m}$ for \textit{A} and \textit{B}
polaritons, respectively. The corresponding effective magnetic
lengths can be estimated as $r_{BA}^{(eff)} = 1.236 \ \mathrm{mm}$
and $r_{BB}^{(eff)} = 1.124 \ \mathrm{mm}$. Therefore, $\xi_{A} \ll
r_{BA}^{(eff)}$ and $\xi_{B} \ll r_{BB}^{(eff)}$ for \textit{A} and
\textit{B} polaritons, respectively, and the Thomas-Fermi
approximation is valid for the system under consideration.

The Bogoliubov approximation for the dilute weakly-interacting Bose
gas of polaritons results in the sound spectrum of collective
excitations at low momenta~\cite{Abrikosov,Griffin}: $\varepsilon
(P)=c_{S}P$ with the sound velocity~\cite{BLS}: $c_{S}=\left(
U_{\mathrm{eff}}^{(0)}n/M_{p}\right) ^{1/2}=\left(
3ke^{2}a_{2D}n/(2\epsilon M_{p})\right) ^{1/2}$. Let us consider the
microcavity polaritons at low temperatures in the presence of
superfluidity when the effective magnetic $\mathbf{B}_{\sigma
}^{(eff)}$ and electric $\mathbf{E}^{(eff)}$ fields are given by Eqs.~(\ref%
{Beff_Aeff}). In the presence of superfluidity, the polariton system
has two components: superfluid and normal~\cite{Abrikosov,Griffin}.
The superfluid-normal phase transition in this 2D system is the
Kosterlitz-Thouless transition~\cite{Kosterlitz}, and the
temperature of this transition $T_{c}$ in a 2D microcavity polariton
system is determined as:
\begin{equation}
T_{c}=\frac{\pi \hbar ^{2}n_{s}(T_{c})}{2k_{B}M_{p}}\ , \label{T_KT}
\end{equation}%
where $n_{s}(T)$ is the concentration for the superfluid component
of the polariton system~\cite{Abrikosov,Griffin} in a microcavity as a function of temperature $T$, and $%
k_{B}$ is the Boltzmann constant.

The polaritons in the superfluid component do not collide, and
therefore, the scattering time of microcavity polaritons is $\tau
\rightarrow +\infty $. In this case, one obtains the transport
equation for microcavity polaritons from Eq.~(\ref{tran}), which can
be rewritten for the \textit{x}- and \textit{y}-components as
\begin{equation}
M_{p}\frac{dv_{x}}{dt}=-\eta _{\sigma }B^{(eff)}v_{y},\hspace{1cm}M_{p}\frac{%
dv_{y}}{dt}=E^{(eff)}+\eta _{\sigma }B^{(eff)}v_{x}.
\label{transupcom}
\end{equation}%
If the initial conditions for Eq.~(\ref{transupcom}) are $v_{0x} =
v_{0y} = 0$, the superfluid polaritons will be accelerated until the
system reaches steady state, which corresponds to $d v_{x}/dt
= d v_{y}/dt = 0$. According to Eq.~(\ref{transupcom}), in the steady state $%
v_{y} = 0$ and $v_{x} = - E^{(eff)}/\eta_{\sigma}B^{(eff)}$.
Defining the linear superfluid polariton flow density as
$\mathbf{j}^{(s)} = n_{s}
\mathbf{v}$, one obtains the conductivity tensor $\tilde{\sigma%
}_{\sigma}^{(s)} (T)$ for the superfluid with the following
components:
\begin{equation}
\sigma_{\sigma xx}^{(s)} = \sigma_{\sigma yy}^{(s)} = 0 ,
\hspace{1cm}
\sigma_{\sigma xy}^{(s)}(T) = -\sigma_{\sigma yx}^{(s)} (T) = - \frac{%
n_{s}(T)}{\eta_{\sigma}B^{(eff)}} \ .  \label{consup}
\end{equation}
For the conductivity tensor $\tilde{\sigma}_{\sigma}^{(n)} (T)$ for
the normal component one can use Eq.~(\ref{conductivity}), substituting $%
\sigma_{0}(T) = \tau n_{n}(T)/M_{p}$, where $n_{n}(T)$ is a 2D
concentration of the normal component~\cite{Abrikosov,Griffin}.
The total conductivity tensor in the presence of superfluidity is given by $%
\tilde{\sigma}_{\sigma}^{(tot)} (T) = \tilde{\sigma}_{\sigma}^{(s)}
(T) + \tilde{\sigma}_{\sigma}^{(n)} (T)$.

Following the procedure~\cite{Abrikosov} we obtain the superfluid
density as $n_{s}(T) = n- n_{n}(T)$ by determining the density of
the normal component $n_{n}(T)$ as a linear response of the total
momentum with respect to the external velocity:
\begin{equation}
\label{n_n} n_{n}(T) = \frac{3 \zeta (3) }{2 \pi \hbar^{2}}
\frac{k_{B}^{3}T^3}{c_S^4 M_{p}} \ .
\end{equation}

Since the diagonal components of the conductivity tensor for the
superfluid component $\sigma_{\sigma xx}^{(s)} = \sigma_{\sigma
yy}^{(s)}$ equal to zero, only the normal component contributes to
the diagonal components of the total conductivity tensor
$\sigma_{\sigma xx}^{(tot)} = \sigma_{\sigma yy}^{(tot)}=
\sigma_{\sigma xx}^{(n)} = \sigma_{\sigma yy}^{(n)}$. In the
presence of superfluidity at $T< T_{c}$ the diagonal components of
the total conductivity tensor are directly proportional to the
concentration of the normal component $n_{n}(T)$. The latter
increases according to Eq.~(\ref{n_n}) as $T^{3}$. Thus one can
determine $n_{n}(T)$ and $n_{s}(T)$ by a measurement of
$\sigma_{\sigma xx}^{(tot)}$ or $\sigma_{\sigma yy}^{(tot)}$ at
different temperatures at $T<T_{c}$. At $T\geq T_{c}$ the
concentration of the normal component equals to the total
concentration of polaritons. Therefore, by the measurement of the
diagonal components of the total conductivity tensor, one can
determine the Kosterlitz-Thouless phase transition temperature
$T_{c}$.

Let us mention that the components of the
conductivity tensor can be obtained via the linear polariton flow  density $%
\mathbf{j}$, which is defined by the polariton flow. The polariton
flow is
determined by the component of the total polariton momentum $\mathbf{P}%
_{\parallel}$ in the direction parallel to the Bragg mirrors. The
component $\mathbf{P}_{\parallel}$ can be calculated from the
experimental measurement of the angular intensity distribution of
the photons escaping the optical microcavity, similar to the
experiment suggested in Ref.~\cite{BKL}. The polariton flow was
obtained experimentally recording directly the momentum distribution
of the particles by angle resolving the
far-field photon emission from the polaritons, because $\mathbf{P}%
_{\parallel}$ has a one-to-one correspondence with the external
angle of photon emission, which was measured~\cite{Snoke_PRX}. Also
the polariton flow can be obtained by using the first order spatial
correlation function for a polariton condensate, which was measured
experimentally by employing a Michelson interferometer
setup~\cite{Yamamoto_PNAS,Caputo}.

Since at the Kosterlitz-Thouless transition temperature $T=T_{c}$
the universal jump in the superfluid concentration
occurs~\cite{Kosterlitz}, one can determine $T_{c}$ by observation
of the jumps at $T=T_{c}$ in  $\sigma_{\sigma xx}^{(tot)} (T)$ and
$\sigma_{\sigma xy}^{(tot)} (T)$ components of the total
conductivity tensor as functions of temperature $T$. This is
possible, because the coefficients of proportionality in the
dependencies of $\sigma_{\sigma xy}^{(n)} (T)$ and $\sigma_{\sigma
xy}^{(s)} (T)$   on $n_{n}(T)$ and $n_{s}(T)$, correspondingly, are different. Besides, only the normal concentration $%
n_{n}(T)$ contributes to  $\sigma_{\sigma xx}^{(n)} (T)$ and
$\sigma_{\sigma yy}^{(n)} (T)$. Also by observation of
$\sigma_{\sigma xx}$ and $\sigma_{\sigma xy}$ one can determine a
scattering time of microcavity polaritons $\tau$ by using
Eq.~(\ref{conductivity}).  Our calculations show that the
contribution to $\sigma_{\sigma xy}^{(tot)} (T)$ and therefore the
Hall linear polariton flow density is mainly given by the superfluid
component, while the contribution from the normal component is
negligible.

Substituting Eq.~(\ref{n_n}) for the density $n_{s}$ of the
superfluid component into Eq.~(\ref{T_KT}), one obtains an equation
for the Kosterlitz-Thouless transition temperature $T_{c}$. The
solution of this equation is
\begin{equation}
T_c = \left[\left( 1 + \sqrt{\frac{32}{27}\left(\frac{M_{p} k_{B}T_{c}^{0}}{%
\pi \hbar^{2} n}\right)^{3} + 1} \right)^{1/3} - \left(
\sqrt{\frac{32}{27} \left(\frac{ M_{p} k_{B}T_{c}^{0}}{\pi \hbar^{2}
n}\right)^{3} + 1} - 1 \right)^{1/3}\right] \frac{T_{c}^{0}}{
2^{1/3}} \ ,  \label{tct}
\end{equation}
where $T_{c}^{0}$ is the temperature, corresponding to vanishing
superfluid density in the mean-field approximation, when
$n_{s}(T_{c}^{0}) = 0$,
\begin{equation}
T_c^0 = \frac{1}{k_{B}} \left( \frac{ \pi \hbar^{2} n c_s^4 M_{p}
}{6 \zeta (3)} \right)^{1/3} \ .  \label{tct0}
\end{equation}

At the temperature $T= 300 \ \mathrm{K}$  for \textit{A} polaritons
we have obtained the total Hall linear polariton flow density in the
presence of superfluidity $j_{x}^{(tot)}= 8.51887 \times 10^{13} \
\mathrm{nm^{-1} s^{-1}}$, $j_{x}^{(tot)}= 9.54342 \times 10^{13} \
\mathrm{nm^{-1} s^{-1}}$, $j_{x}^{(tot)}= 9.76334 \times 10^{13} \
\mathrm{nm^{-1} s^{-1}}$, $j_{x}^{(tot)}= 1.1415 \times 10^{14} \
\mathrm{nm^{-1} s^{-1}}$ for $\mathrm{W Se_{2}}$ for $\mathrm{Mo
S_{2}}$, $\mathrm{Mo Se_{2}}$, and $\mathrm{W Se_{2}}$,
respectively. At the temperature $T = 300 \ \mathrm{K}$ for
\textit{B} polaritons we have obtained the total Hall linear
polariton flow density in the presence of superfluidity
$j_{x}^{(tot)}=1.22581 \times 10^{14} \ \mathrm{nm^{-1} s^{-1}}$,
$j_{x}^{(tot)}= 1.32831 \times 10^{14} \ \mathrm{nm^{-1} s^{-1}}$,
$j_{x}^{(tot)}= 1.09252 \times 10^{14} \ \mathrm{nm^{-1} s^{-1}}$,
$j_{x}^{(tot)}= 1.19047 \times 10^{14} \ \mathrm{nm^{-1} s^{-1}}$
for $\mathrm{Mo S_{2}}$, $\mathrm{Mo Se_{2}}$, and $\mathrm{W
Se_{2}}$, respectively. Therefore, for \textit{A} polaritons the
total Hall linear polariton flow density is the largest for   a
$\mathrm{W Se_{2}}$ monolayer and the smallest for a  $\mathrm{Mo
S_{2}}$ monolayer, while for \textit{B} polaritons it is the largest
for a $\mathrm{Mo Se_{2}}$ monolayer and the smallest for a
$\mathrm{W S_{2}}$ monolayer. Also, $j_{x}^{(tot)}$ for \textit{B}
polaritons is larger than for \textit{A} polaritons for the same
monolayer.

Let us mention that the results, presented in this Section, are
applicable for \textit{A} and \textit{B} polaritons. \textit{A}
polaritons are formed via \textit{A} excitons coupled to microcavity
photons, and \textit{B} polaritons are formed via \textit{B}
excitons coupled to microcavity photons, correspondingly. The
effective masses of \textit{A} and \textit{B} excitons in a TMDC
heterostructure are given by $M_{A}=m_{e\uparrow }+m_{h\downarrow }$ and $%
M_{B}=m_{e\downarrow }+m_{h\uparrow }$, where $m_{e\uparrow
(\downarrow )}$ is the effective mass of spin-up (spin-down)
electrons from the conduction band and $m_{h\uparrow (\downarrow )}$
is the effective mass of spin-up (spin-down) holes from the valence
band, correspondingly. In our formulas
above we assume that for \textit{A} and \textit{B} excitons mass $M$ should be replaced by $%
M_{A}$ and $M_{B}$, respectively. Correspondingly, the polariton
effective mass $M_{p}$ should be replaced by $M_{pA}$ and $M_{pB}$
for \textit{A} and \textit{B} polaritons, respectively.

\section{On the observation of SHEP}

\label{exper}

The observation of the SHEP in TMDC is related to the measurement of
the shift of the angular distribution of the photons escaping the
optical microcavity due to the effective gauge magnetic and electric
fields acting on polaritons. In the absence of the effective gauge
magnetic and electric fields, the angular distribution of the
photons escaping the microcavity is central-symmetric with respect
to the perpendicular to the Bragg mirrors. In order to analyze the
deflection of the polariton flow in the $(x,y)$ plane of the
microcavity due to the SHEP one can measure the average tangent of
the angle $\alpha$ of deflection for the polariton flow, defined as
\begin{equation}
\overline{\tan\ \alpha}= \left| \frac{j_{x}}{j_{y}} \right| =
\left|\frac{\sigma_{\sigma xy}}{\sigma_{\sigma yy}} \right| . \label{tanxy}%
\end{equation}
At the absence of superfluidity by substituting
Eq.~(\ref{conductivity}) into Eq.~(\ref{tanxy}) one obtains
$\overline{\tan\ \alpha} = \omega_{c} \tau$. The same expression is
valid in the presence of superfluidity for the angle of deflection
of the flow of the normal component, since
\begin{equation}
\overline{\tan\ \alpha^{(n)}}= \left|
\frac{j_{x}^{(n)}}{j_{y}^{(n)}} \right| = \left|\frac{\sigma_{\sigma
xy}^{(n)}}{\sigma_{\sigma yy}^{(n)}} \right| = \omega_{c} \tau  . \label{tanxyn}%
\end{equation}
Substituting $l = a^{2}/8y_{0}$, $a = 10 \ \mathrm{\mu m}$, $y_{0} =
2.5 \ \mathrm{\mu m}$, $|k_{1}|+ |k_{2}| \approx 3 \ \mathrm{\mu
m}^{-1}$, $\tau = 1.040 \ \mathrm{ps}$ into Eq.~(\ref{tanxyn}), one
can estimate for $\mathrm{WSe_{2}}$, $\overline{\tan\ \alpha^{(n)}}
\approx 10^{-5}$, and $\overline{\alpha^{(s)}} \approx 10^{-3 \ 0}$.
Therefore, the normal component of the polariton system almost does
not deflect in the direction perpendicular to the effective gauge
electric field due to the SHEP.

The average tangent of the angle $\alpha^{(s)}$ of deflection of the
flow of the superfluid component can be obtained as
\begin{equation}
\overline{\tan\ \alpha^{(s)}}= \left|
\frac{j_{x}^{(s)}}{j_{y}^{(s)}} \right| = \left|\frac{\sigma_{\sigma
xy}^{(s)}}{\sigma_{\sigma yy}^{(s)}}
\right| \rightarrow + \infty , \label{tanxys}%
\end{equation}
because $\sigma_{\sigma yy}^{(s)} = 0$ according to
Eq.~(\ref{consup}). Therefore, one has $\overline{\alpha^{(s)}} =
90^{0}$ for the average angle of deflection of the superfluid
component due to the SHEP.

Since the cyclotron frequency $\omega_{c}$ for polaritons is the
same by the order of magnitude as for excitons $\omega_{c} \sim 1/M$
and a scattering time of microcavity polaritons $\tau \sim 10^{-12}
\ \mathrm{s}$, the deflection of the normal component of polariton
flow is very small, but is different for \textit{A} and \textit{B}
polaritons. In contract, as one can see from Eqs.~(\ref{consup})
and~(\ref{tanxys}) that $\overline{\alpha^{(s)}}\rightarrow 90^{0}$
and is the same for all different TMDC monolayers embedded in
different microcavities. Therefore, due to the SHEP the \textit{A}
and \textit{B } polariton flows are splitting. The normal components
of the \textit{A} and \textit{B} polariton flows are slightly
deflected in opposite directions and propagate almost
perpendicularly to the counterpropagating beams. In contrast, the
superfluid components of the \textit{A} and \textit{B} polariton
flows are propagated in opposite directions along the
counterpropagating beams. Thus, one can separate the superfluid and
normal components of the polariton flow. Therefore, by measuring the
angular distribution of the photons escaping the optical microcavity
in the presence and absence of the effective gauge magnetic and
electric fields and determining the shift of these distributions,
one can observe the SHEP.

Note that the cause of the shift of the angular distribution of the
photons escaping from the superfluid component due to the SHEP is
different from the reason of the shift of the angular distribution
of the photons escaping from the normal component in the polariton
drag effect~\cite{BKL,BKL_PLA}. This difference is related to the
fact that the drag effect is caused only by the excitations, while
the gauge fields can influence the entire superfluid component.

\section{Possible technological applications of SHEP}

\label{tech}

We propose the optical switch based on microcavity polaritons,
formed by excitons in a TMDC monolayer, in the SHEP regime. We
propose the switch for a polariton flow. We consider the polariton
system in the presence of superfluidity. Using circular polarized
pumping, one can excite both \textit{A} and \textit{B} excitons in
one valley simultaneously. In this case, \textit{A} and \textit{B}
polaritons are formed due to coupling of \textit{A} and \textit{B}
excitons to the microcavity photons, correspondingly.
 Due to the SHEP, there will be two
different by magnitude superfluid spin Hall polariton
 flows perpendicular to the effective gauge electric field in opposite directions for \textit{A} and
\textit{B} polaritons. We can control these two different
 superfluid polariton flows by changing the
concentration of excited \textit{A} and/or \textit{B} polaritons. A
polariton flow occurs also in the direction along  the effective
gauge electric field.

Besides, one can excite either \textit{A} or \textit{B} polaritons
once at the time. We can switch the magnitude and direction of
superfluid polariton  Hall  flow perpendicular to the effective
gauge electric field by switching the frequency of laser pumping,
exciting either \textit{A} or \textit{B} polaritons. By switching
from the regime of \textit{A} excitons (or \textit{A} polaritons) to
\textit{B} excitons (or \textit{B} polaritons) or vise versa, we can
switch the direction of superfluid polariton Hall flow to the
opposite one and change the magnitude of superfluid polariton Hall
flow. However, the magnitude of this superfluid polariton flow will
be different for \textit{A} and \textit{B} polaritons. This
magnitude of the superfluid polariton flow can be controlled by
exciting either \textit{A} or \textit{B} polaritons.

\section{Discussion}

\label{disc}

Another important feature of the SHEP in TMDC is that the conductivity tensor $\tilde{%
\sigma}$ depends on the exciton mass $M,$ which is different for
\textit{A} and \textit{B} excitons. Therefore, the conductivities
and polariton flows (including Hall conductivities and Hall flows)
will have different magnitudes for \textit{A} and \textit{B}
polaritons. Besides, the Hall polariton flows have different
directions for \textit{A} and \textit{B} polaritons, since they
depend on the spin orientation factor $\eta_{\sigma}$. The SHEP in
TMDC  differs for \textit{A} and \textit{B} polaritons by twofold:
i. the polaritonic flow has different magnitudes due to the
different masses of \textit{A} and \textit{B} polaritons; ii. the
Hall polariton flows will have different directions for \textit{A}
and \textit{B} polaritons, since they depend on the spin orientation
factor.

 The SHEP can be observed for both
linear and circular polarized light pumping. For linear polarized
light the excitons, forming polaritons, are created in both valleys
$\mathbf{K}$ and $\mathbf{-K}$. In this case \textit{A} polaritons
from  valley $\mathbf{K}$ and \textit{B} polaritons from the
$\mathbf{-K}$ valley will be deflected in one direction, while the
\textit{B} polaritons from the valley $\mathbf{K}$ and \textit{A}
polaritons from the
 valley $\mathbf{-K}$  will be deflected in opposite
direction. For left circular polarized light \textit{A} and
\textit{B} polaritons from the valley $\mathbf{K}$ will be deflected
in opposite directions. Analogously, for right circular polarized
light \textit{A} and \textit{B} polaritons from the valley
$\mathbf{-K}$ will be deflected in opposite directions.

 Let us mention that the EHE in
monolayer MoS$_{2}$ and valley-selective spatial transport of
excitons on a micrometre scale were directly observed by
polarization-resolved photoluminescence mapping~\cite{Onga}. This
EHE studied in Ref.~\cite{Onga} is caused by intrinsic properties of
a TMDC material. While the EHE is very important for valleytronics,
the proposed SHEP is caused by the properties of two external
counterpropagating laser beams, and, therefore, can be controlled by
changing the parameters of these laser beams. The study the
possibilities of application of the EHE~\cite{Onga} for microcavity
polaritons will be performed in subsequent work.

 Let us discuss at this point the
difference of the SHEP and OSHEP~\cite{OSH1,OSH2}. The  SHEP and the
OSHEP~\cite{OSH1,OSH2} have two cardinally different mechanisms. In
the OSHEP the deflection of polaritons appears due to the deflection
of the photon component of polaritons. The OSHEP was analyzed for
polaritons, formed by excitons in a GaAs quantum well, embedded in a
high-quality GaAs/AlGaAs microcavity~\cite{OSH2}.  The OSHEP is
controlled by the linear polarization of the laser
pump~\cite{OSH1,OSH2}. In contrast, in the SHEP in TMDC the
deflection of polaritons appears due to the deflection of the
exciton component of polaritons.  The SHEP in TMDC appears due
coupling of two spatially varying infrared laser beams to the
internal levels of the excitons in TMDC, considered for excitons in
Ref.~\cite{li}.  If under linear polarized pumping required for
OSHEP~\cite{OSH1,OSH2}, both valleys would be populated, the
polaritons corresponding to the, for example, $\mathbf{K}$ valley
\textit{A} excitons and $\mathbf{-K}$ valley \textit{A} excitons
would deflect to different edges of cavity. In contract, for the
SHEP in TMDC the circular polarized pumping, which creates excitons
in only one chosen valley, allows to initiate separated Hall
polariton flows of \textit{A} and \textit{B} polaritons from the
same valley in opposite directions.
The latter allows to control the spin Hall effect in a chosen
valley, which is very useful for valleytronics. Since elastic
scattering of photons by disorder is the main scattering mechanism
required for the OSHEP~\cite{OSH1,OSH2}, in the case of weak
disorder in TMDC the SHEP proposed in our paper is the dominant
effect, because the SHEP is not caused by disorder. Moreover, since
the SHEP can be observed for both  circular and linear polarized
light pumping, in the latter case one can modulate the signals from
two counterpropagating laser beams either by changing periodically
the distance $l$ between two laser beams or by a rotating screen
with a hole  in front of the source of two laser beams. In this
case, when the signals from two laser beams are off, the SHEP is
absent and only the OSHEP is present, while in the presence of the
signals from two laser beams the both SHEP and OSHEP are present.
The latter allows one to analyze the contribution from SHEP  to the
defected flow of photons, escaping the microcavity. By this way the
SHEP can be registered as the modulated component of the signal.
Another essential difference between the SHEP and OSHEP is that the
SHEP provides the deflection of polariton superfluid component,
while the OSHEP does not result in deflection of the polariton
superfluid component due to the absence of the scattering of
superfluid component by disorder.

The novel 2D materials such as transition metal dichalcogenides
~\cite{Kormanyos}, germanene~\cite{Tabert,Lu}, and
stanene~\cite{Saxena} are characterized by relatively large exciton
binding energies and strong spin-orbit coupling. However, in
contrast to TMDC, germanene and stanene demonstrate high exciton
binding energies and strong SOC only under strong perpendicular
electric field. We assume that the SHEP can occur for microcavity
polaritons, formed by excitons only in the novel 2D materials such
as either TMDCs or germanene and stanene under high perpendicular
electric field due to strong SOC. In contrast the SHEP cannot be
observed for microcavity polaritons, formed by excitons in a
semiconductor quantum well due to absence of strong SOC. Let us
mention that while we study exciton polaritons formed by excitons in
TMDCs, our approach seems to be applicable for all novel 2D
materials with strong SOC, including germanene and stanene under
high electric field.

 Let us emphasize the importance of considered in this
paper spin Hall effect for polaritons. By using the SHEP, one can
control the flows of photons,  and induced by SHEP the flows of
polaritons lead to flows of photons, escaping the microcavity. So we
suggest the method to control photon flows. Another advantage of
consideration of SHEP is the possibility of observation of spin Hall
effect in the superfluid system.

\section{Conclusions}

\label{conc}

We have proposed the spin Hall effect for microcavity polaritons,
formed by excitons in a TMDC and microcavity photons. We
demonstrated that the polariton flow can be achieved by generation
the effective gauge vector and scalar potentials, acting on
polaritons.
We have obtained the components of polariton conductivity tensor for
both non-interacting polaritons without BEC and for
weakly-interacting Bose gas of polaritons in the presence of BEC and
superfluidity. These results for non-interacting polaritons are
applicable for the two-component system of \textit{A} and \textit{B}
polaritons. We have studied the SHEP for both superfluid and normal
components.  Let us emphasize that induced by SHEP the flows of
polaritons lead to flows of photons, escaping the microcavity.
Therefore, one can control the flows of photons. Another advantage
of of SHEP is the possibility of observation of spin Hall effect in
the superfluid system. For circular polarized light we demonstrated
that due to the SHEP the polariton flows in the same valley are
splitting: the normal component of the \textit{A} and \textit{B}
polariton flows slightly deflect in opposite directions and
propagate almost perpendicularly to the counterpropagating beams,
while the superfluid components of the \textit{A} and \textit{B}
polariton flows propagate in opposite directions along the
counterpropagating beams. Thus, one can separate the superfluid and
normal components of the polariton flow.  Since only the superfluid
component contributes to the polariton Hall flow, while the normal
component contributes only to the flow almost parallel to the
effective gauge electric field, the SHEP can be employed to separate
the superfluid component from the normal component. For the linear
polarized light one can observe the same effect for the \textit{A}
and \textit{B} polariton flows in $\mathbf{K}$ and $\mathbf{-K}$
valleys, respectively. Observation of the SHEP in the presence of
superfluidity can be achieved by measuring the angles $\alpha^{(n)}$
and $\alpha^{(s)}$ of deflections for the normal and superfluid
polariton flows.

\acknowledgments

O.~L.~B. and R.~Ya.~K. are supported by US Department of Defense
under Grant No. W911NF1810433 and PSC CUNY under Grant No. 60599-00
48. Yu.~E.~L. is supported by Program of Basic Research of National
Research University HSE and by the RFBR Grant (17-02-01134).



\appendix

\section{The Hamiltonian for a TMDC exciton coupled to two
counterpropagating Gaussian laser beams}

\label{app:C}

We consider two infrared laser beams linearly polarized along the
$y$ axis, which propagate along the $x$ axis, acting on the TMDC
excitons, forming polaritons. The exciton ground state $|g\rangle
\equiv |1s\rangle$ and two low excited states $|1\rangle \equiv |2
p_{y}\rangle$ and $|2\rangle \equiv |3 p_{y}\rangle$ were under
consideration in Ref.~\cite{li}. These two laser beams couple
$|g\rangle$ to $|1\rangle$ and $|2\rangle$, correspondingly, with
equal detuning $\delta$. Two counterpropagating Gaussian laser beams
are characterized by symmetrical centers, shifted along the $y$
axis,
and spatial profiles $e E_{1(2)}(\mathbf{R}) \langle g|\mathbf{e}_{y}\cdot%
\mathbf{r}|1(2)\rangle/\hbar = \Omega_{1(2)}(\mathbf{R})e^{i\phi_{1(2)}(%
\mathbf{R})}$, where $e$ is the electron charge,
$E_{1(2)}(\mathbf{R})$ is the external electric field, $\mathbf{R} =
(x,y)$ is the coordinate vector
of the center-of-mass of an exciton, $\Omega_{1} = \Omega_{0}\exp\left[%
-(y-y_{1})^{2}/a^{2}\right]$ and $\Omega_{2} = \Omega_{0}\exp\left[%
-(y-y_{2})^{2}/a^{2}\right]$ are  Rabi frequencies that
characterized the beams, $y_{1} = -y_{2}=y_{0}$, $\phi_{1}
(\mathbf{R}) = k_{1}x$, $\phi_{2} (\mathbf{R}) = k_{2}x$.

The effective center-of-mass Hamiltonian $H_{\sigma}$ for an
exciton, associated with different spin states of the conduction
band electron, forming an exciton, $\sigma = \uparrow$ and $\sigma =
\downarrow$,  in a
TMDC coupled to two coordinate dependent laser beams is given by~%
\cite{li}
\begin{equation}
H_{\sigma} = \frac{1}{2M} \left(\mathbf{P} -
\mathbf{A}_{\sigma}\right)^{2} + V_{\sigma} ,  \label{Hamed}
\end{equation}
where $\mathbf{P}$ is the momentum of the center-of-mass of an
exciton, $M = m_{e} + m_{h}$ is the exciton total mass ($m_{e}$ and
$m_{h}$ are the effective masses of an
electron and a hole in TMDC, respectively, $V_{\sigma}$ and $\mathbf{}%
_{\sigma}$ are the spin-dependent gauge scalar and vector potential,
correspondingly, which depend on $\mathbf{R}$. The effective gauge
magnetic field $\mathbf{B}_{\sigma}$ is defined as
$\mathbf{B}_{\sigma} = \nabla_{\mathbf{R}} \times
\mathbf{A}_{\sigma}$. Two counterpropagating Gaussian laser beams
with the centers, shifted along the $y$ axis, produce a coordinate
dependent gauge field $\mathbf{B}_{\sigma}$~\cite{li}:
\begin{equation}
\mathbf{A}_{\sigma} = \frac{\eta_{\sigma}\hbar\left(|k_{1}|+ |k_{2}|\right)}{%
1 + e^{-y/l}}\mathbf{e}_{x}, \ \ \ \ \ \mathbf{B}_{\sigma} = \frac{%
-\eta_{\sigma}\hbar\left(|k_{1}|+ |k_{2}|\right)}{4l\cosh^{2}(y/2l)}\mathbf{e%
}_{z},  \label{AB}
\end{equation}
where  $\eta_{\uparrow} = 1$  and $\eta_{\downarrow} = -1$ for
\textit{A} and \textit{B} excitons, respectively, $l =
a^{2}/8y_{0}$, $a = 10 \ \mathrm{\mu m}$ is the beam width, $y_{0} =
2.5 \ \mathrm{\mu m}$ is the spatial shift of two laser beams,
$|k_{1}|+ |k_{2}| \approx 3 \ \mathrm{\mu m}^{-1}$.

Below we provide the qualitative analysis for the validity of the
assumption
that the exciton gauge magnetic field $\mathbf{B}_{\sigma}(y)$ given by Eq.~(%
\ref{ABexp}) and scalar potential $V_{\sigma} (y)$ can be treated as
coordinate independent constants. The aforementioned assumption is
valid, if $y$ is small compared with $l$ ($l = 5 \ \mathrm{\mu
m}$~\cite{li}) and $y \ll \tilde{y}(y)$, where $\tilde{y}(y)$ is the
characteristic length of changes in the exciton gauge magnetic field
and scalar potential, defined as
\begin{equation}
\tilde{y}(y) = \left|\frac{B(y)}{d B(y)/d y} \right| = \left|\frac{V_{\sigma}%
}{d V_{\sigma} (y)/d y} \right| = \frac{l}{\tanh\left(\frac{y}{2
l}\right)}, \label{char_le}
\end{equation}
where $B(y) = \left|\mathbf{B}_{\sigma}(y)\right|$. Assuming that
$y$ does not exceed $2.5 \ \mathrm{\mu m}$~\cite{li}, one obtains
$\tilde{y} (y) \geq \tilde{y} \left(2.5 \ \mathrm{\mu m}\right) =
20.41 \ \mathrm{\mu m}$. Therefore, since at $y \leq 2.5 \
\mathrm{\mu m}$ the inequality $y \ll \tilde{y}(y)$ together with
the assumption about small $y$ compared with $l$
hold, we assume that in our system the exciton gauge magnetic field $\mathbf{%
B}_{\sigma}(y)$ given by Eq.~(\ref{ABexp}) and scalar potential $%
V_{\sigma}(y)$ do not depend on coordinates and, therefore, are
constants.

Assuming $y/l \ll 1$ at $y \ll 5 \ \mathrm{\mu m}$, we expand
$\mathbf{A}_{\sigma}$ and $\mathbf{B}_{\sigma}$ in series in terms
of  $y/l$ and in the first order approximation obtain from
Eq.~(\ref{AB}) the following:
\begin{equation}
\mathbf{A}_{\sigma} = \frac{\eta_{\sigma}\hbar\left(|k_{1}|+ |k_{2}|\right)}{%
2}\left(1 + \frac{y}{2l}\right)\mathbf{e}_{x}, \ \ \ \ \
\mathbf{B}_{\sigma} = \frac{-\eta_{\sigma}\hbar\left(|k_{1}|+
|k_{2}|\right)}{4l}\mathbf{e}_{z}, \label{ABexp}
\end{equation}

The spin-dependent gauge scalar potential $V_{\sigma}$ is given
by~\cite{li}:
\begin{equation}
V_{\sigma} (\mathbf{R}) = \lambda_{\sigma} + W (\mathbf{R}) ,  \label{Vs}
\end{equation}
where
\begin{equation}
\lambda_{\downarrow} = \hbar \delta , \ \ \ \ \ \lambda_{\uparrow} \approx
\hbar \delta + \frac{\hbar \Omega^{2}}{4 \delta} ,  \label{lam}
\end{equation}
and $\Omega \equiv \left(\Omega_{1}^{2} +
\Omega_{2}^{2}\right)^{1/2}$. Then one obtains from Eq.~(\ref{lam})
the following expression:
\begin{equation}
\lambda_{\uparrow} \approx \hbar \delta + \frac{\hbar
\Omega_{0}^{2}}{2 \delta}
e^{-2y_{0}^{2}/a^{2}}e^{-2y^{2}/a^{2}}\cosh(y/2l).  \label{lam2}
\end{equation}
In Eq.~(\ref{Vs}), the scalar potential $W (\mathbf{R})$ is given by
\begin{equation}  \label{WR}
W (\mathbf{R}) = \frac{\hbar^{2}}{2M} \left(\left|\nabla_{\mathbf{R}%
}\theta\right|^{2} + \sin^{2}\theta \cos^{2}\theta \left|\nabla_{\mathbf{R}%
}\phi (\mathbf{R})\right|^{2}\right),
\end{equation}
where $\theta = \tan^{-1}\left(\Omega_{1}/\Omega_{2}\right)$, $\phi (\mathbf{%
R})\equiv \phi_{1} (\mathbf{R}) - \phi_{2} (\mathbf{R})$. From Eq.~(\ref{WR}%
) the following expression can be derived
\begin{equation}  \label{WR2}
W (\mathbf{R}) = \frac{\hbar^{2}}{2M} \left(\frac{4y_{0}^{2}}{a^{4}} + \frac{%
\left(|k_{1}|+ |k_{2}|\right)^{2}}{4} \right) \frac{1}{\cosh^{2}(y/2l)} .
\end{equation}
Since we consider two counterpropagating Gaussian laser beams with
symmetrically shifted centers along the $y$ axis~\cite{li}, we used
for derivation of Eq.~(\ref{WR2}), the following relation: $|k_{1} -
k_{2}| = |k_{1}|+ |k_{2}|$. Assuming $y/l \ll 1$ at $y \ll 5 \
\mathrm{\mu m}$, in the first order with respect to $y/l$, one
obtains $W = \mathrm{const}$. Therefore, in our approach we will not
consider the gauge scalar potential acting on excitons, since in the
first order with respect to $y/l$ it results in zero scalar gauge
field, because the nonzero scalar potential occurs only in the
second order on $y/l$.

\section{Diagonalization of the Hamiltonian $\hat{\mathcal{H}}$ by using unitary
transformation}

\label{app:B}

We apply the quasilocal approximation   which  can be used for the momenta $%
P$, obeying to the condition $Pr_{B}^{(eff)}\gg\hbar$, where
$r_{B}^{(eff)} = \sqrt{\hbar/B^{(eff)}}$ is the effective magnetic
length, and $B^{(eff)}$ is the magnitude of the effective magnetic
field, acting on polaritons, defined by Eq.~(\ref{Beff_Aeff}). In
this quasiclassical approach the coordinate $y$, entering the
exciton energy dispersion $\varepsilon_{ex}(\mathbf{P})$ through the
gauge vector potential $\mathbf{A}_{\sigma}$, is considered to be a
number parameter rather than  an operator.

If one assumes $\hat{H}_{exc-exc}=0$, the Hamiltonian
$\hat{\mathcal{H}}$ can be diagonalized by using unitary
transformation and can be written as~\cite{Ciuti}:
\begin{equation}  \label{h0}
\hat{H}_{0} = \sum_{\mathbf{P}}\varepsilon_{LP}(\mathbf{P})\hat{p}_{\mathbf{P%
}}^{\dagger}\hat{p}_{\mathbf{P}} +\sum_{\mathbf{P}}\varepsilon_{UP}(\mathbf{P%
})\hat{u}_{\mathbf{P}}^{\dagger}\hat{u}_{\mathbf{P}},
\end{equation}
where $\hat{p}_{\mathbf{P}}^{\dagger}$ and
$\hat{u}_{\mathbf{P}}^{\dagger}$, and $\hat{p}_{\mathbf{P}}$ and
$\hat{u}_{\mathbf{P}}$ are the Bose creation and annihilation
operators for the lower and upper polaritons, correspondingly. The
energy spectra of the lower and upper polaritons are given by
\begin{equation}  \label{eps0}
\varepsilon_{LP/UP}(\mathbf{P}) = \frac{\varepsilon _{ph}(P)
+\varepsilon _{ex}(\mathbf{P})}{2} \mp
\frac{1}{2}\sqrt{(\varepsilon_{ph}(P) - \varepsilon
_{ex}(\mathbf{P}))^{2} + 4|\hbar\Omega_{R}|^{2}},
\end{equation}
where the Rabi splitting between the upper and lower states at $P=0$ equals $%
2\Omega_R$.

The operators of excitons and photons are defined as~\cite{Ciuti}
\begin{equation}  \label{bog_tr}
\hat{b}_{\mathbf{P}} = X_{\mathbf{P}}\hat{p}_{\mathbf{P}} - C_{\mathbf{P}}\hat{u}_{\mathbf{P}%
}, \hspace{3cm} \hat{a}_{\mathbf{P}} = C_{\mathbf{P}}\hat{p}_{\mathbf{P}} + X_{\mathbf{P}}\hat{%
u}_{\mathbf{P}},
\end{equation}
where $X_{\mathbf{P}}$ and $C_{\mathbf{P}}$ are~\cite{Ciuti}
\begin{equation}  \label{bog}
X_{\mathbf{P}} = \frac{1}{\sqrt{1 + \left(\frac{\hbar\Omega_{R}}{\varepsilon_{LP}(%
\mathbf{P}) - \varepsilon _{ph}(P)}\right)^{2}}} , \hspace{3cm}
C_{\mathbf{P}} = - \frac{1}{\sqrt{1 +
\left(\frac{\varepsilon_{LP}(\mathbf{P}) - \varepsilon
_{ph}(P)}{\hbar\Omega_{R}}\right)^{2}}} ,
\end{equation}
and $|X_{\mathbf{P}}|^{2}$ and $|C_{\mathbf{P}}|^{2} = 1 -
|X_{\mathbf{P}}|^{2}$ point out the exciton and cavity photon
fractions in the lower polariton~\cite{Ciuti}.

At $\alpha \equiv 1/2(M^{-1}+(c/%
\tilde{n})L_{C}/\hbar \pi q)P^{2}/|\hbar \Omega _{R}|\ll 1$ we
obtain from Eq.~(\ref{bog}) the following relation: $X_{\mathbf{P}}
\approx 1/\sqrt{2}$.

\end{document}